\documentclass{v16nufact}

\usepackage{subfig}

\def\be{\begin{equation}}
\def\ee{\end{equation}}
\def\bea{\begin{eqnarray}}
\def\eea{\end{eqnarray}}

\newcommand{\Photo}{\includegraphics[height=20mm]{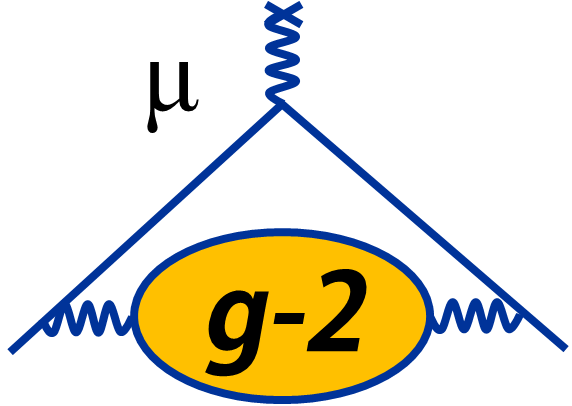}}

\begin{document}
\vspace*{4cm}
\title{The MUON g-2 EXPERIMENT AT FERMILAB}

\author{W. GOHN, \\for the Muon g-2 Collaboration}

\address{Department of Physics and Astronomy, University of Kentucky, 505 Rose St,\\
Lexington, KY 40506, USA}

\maketitle\abstracts{A new measurement of the anomalous magnetic moment of the muon, $a_{\mu} \equiv (g-2)/2$, will be performed at the Fermi National Accelerator Laboratory with data taking beginning in 2017. The most recent measurement, performed at Brookhaven National Laboratory and completed in 2001, shows a 3.5 standard deviation discrepancy with the standard model prediction of $a_\mu$. The new measurement will accumulate 21 times those statistics using upgraded detection and storage ring systems, enabling a measurement of $a_\mu$ to 140 ppb, a factor of 4 improvement in the uncertainty the previous measurement. This improvement in precision, combined with recent and ongoing improvements in the evaluation of the QCD contributions to the $a_\mu$, could provide a 7.5$\sigma$ discrepancy from the standard model if the current difference between experiment and theory is confirmed, a possible indication of new physics. 
}

\section{Introduction}

The currently most precise measurement of the anomalous magnetic moment of the muon $a_\mu$ shows a 3.6 standard deviation discrepancy from the standard model. The measurement was performed by the Brookhaven experiment E821, the final results of which were published in 2006~\cite{Bennett:2006fi}. Is this an indication of new physics beyond the standard model? The discrepancy demonstrated by E821 provides a tantalizing hint at new physics, but lacks the precision to make a definitive case. A new experiment at Fermilab will measure $a_\mu$ with a factor of 21 times the statistics and comparable reduction in systematic uncertainties, which will improve the precision by a factor of four and could increase the discrepancy between the measured and theoretical values to 7.5$\sigma$ if the central value remains unchanged. 

In the Dirac quantum theory~\cite{Dirac:1928hu}, the magnetic moment of a pointlike, spin $1/2$ particle is given by
\begin{equation}
\vec \mu = g\frac{Qe}{2m}\vec s
\label{eq:mu}
\end{equation}
where the $g$ factor in Eq.~\ref{eq:mu} is exactly equal to 2.

\begin{figure}[ht]
\centering
\includegraphics[width=5cm]{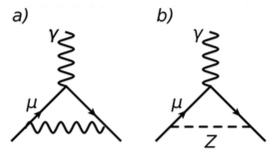}
\includegraphics[width=5cm]{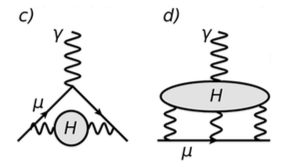}
\caption{Examples of four contributions to $a_\mu$. a) QED Schwinger term; b) Electroweak Z exchange; c) Lowest-order hadronic vacuum polarization; d) Hadronic light-by-light scattering.}
\label{fig:feynman}
\end{figure}

 Since a measurement of the hyperfine structure of atomic hydrogen in 1947~\cite{Nafe:1947zz}, it has been known that g differs from 2 due to other effects, which were later understood to include contributions from quantum electrodynamics, electroweak theory, and  QCD, examples of which are shown in Fig.~\ref{fig:feynman}. Each of these contributions must be computed to the highest possible precision in order to make a valid comparison with the new experimental values~\cite{Blum:2013xva}. If a discrepancy with the standard model value is found, beyond standard model contributions to $g$-$2$ could come from SUSY, dark photons, extra dimensions, or other new physics (NP), as is represented in Eq.~\ref{eq:sum}
\begin{equation}
a_{\mu} = a_{\mu}^{QED} + a_{\mu}^{EW} +a_{\mu}^{QCD} + a_{\mu}^{NP},
\label{eq:sum}
\end{equation}
with the scale of each contribution given by Eq.~\ref{eq:scale}
\begin{equation}
g_{SM} = 2_{Dirac} + \mathcal{O}(10^{-3})_{QED} + \mathcal{O}(10^{-9})_{EW} + \mathcal{O}(10^{-7})_{QCD},
\label{eq:scale}
\end{equation}
and detailed values are shown in Table~\ref{tab:theory}.

The leading contribution to $a_\mu$, and the contribution with the smallest computed uncertainty, is quantum electrodynamics. Recent calculations to 5th order in $\alpha$ have reduced the QED uncertainty to $0.08\times10^{-11}$~\cite{Aoyama:2012wk}. Measurement of the Higgs mass at the LHC further reduces the electroweak uncertainty from $2\times10^{-11}$ to $1\times10^{-11}$~\cite{Gnendiger:2013pva}. 

The largest source of theoretical uncertainty comes from the hadronic contribution to $a_\mu$~\cite{Davier:2010nc}~\cite{Hagiwara:2011af}. There are two major sources of hadronic contributions; hadronic vacuum polarization (HVP) and hadronic light-by-light (HLbL). The leading-order vacuum polarization contribution has traditionally been extracted from measurements of $e^{+}e^{-}\rightarrow hadrons$, which have been performed at BESIII~\cite{Ablikim:2015orh}, BaBar~\cite{Lees:2012cj}, and KLOE~\cite{Babusci:2012rp}. Recently, calculations of the hadronic vacuum polarization from lattice QCD have begun to challenge the scattering experiments for having the smallest theoretical uncertainty on this contribution. Significant effort is being made to calculate both the HPV and HLbL terms on the lattice. Recent results of the disconnected contribution calculated $a_{\mu}^{HPV,LO} = -9.6(3.3)(2.3)\times 10^{-10}$~\cite{Blum:2015you}, and the strange quark connected contribution has been calculated to 2\%, $a_{\mu}^{had}=53.1(9)(_{-3}^{+1})\times 10^{-10}$~\cite{Blum:2016xpd}. Uncertainties match those given by the experimental measurements, but sub-percent precision will require the inclusion of QED and isospin-breaking effects.


The HLbL contribution is smaller, $\mathcal{O}(\alpha^{3})$ as opposed to $\mathcal{O}(\alpha^{2})$. It has recently been calculated for a nearly real pion mass, with a contribution to g-2 of $(132.1 \pm 6.8) \times 10^{-11}$ (with statistical errors only)~\cite{Blum:2015gfa}. Better precision for this calculation requires more computing time, which is underway.

\begin{table}
\centering
\begin{tabular}{|l|c|c|}
\hline
Contribution & $a_\mu$ ($\times 10^{-11}$) &$\delta(a_{\mu})$ ($\times 10^{-11}$)\\
\hline
QED (5 loops)& 116 584 718.95 & $\pm$ 0.08 \\
QCD: HVP (lo) & 6 923 & $\pm$ 42\\
QCD: HVP (ho) & -98.4 & $\pm$ 0.6\\
QCD: HLbL & 105 & $\pm$ 26\\
EW & 154 & $\pm$ 1 \\
\hline
Total SM & 116 591 815 & $\pm$ 49 \\
\hline
\end{tabular}
\caption{Comparison of current standard model theoretical contributions to $a_\mu$.}
\label{tab:theory}
\end{table}

The new experiment is measuring $a_\mu$ at the $\mathcal{O}(10^{-10})$ level, at which several flavors of new physics have the possibility of contributing. All standard model particles contribute to the anomalous magnetic moments of the electron, muon, and tau via vacuum fluctuations. Compared to electrons, the influence of higher mass particles have a higher impact on $a_{\mu}$ by a factor of $(m_{\mu}/m_{e})^2 \approx 4\times10^4$. If the discrepancy from the Standard Model holds, it could be accounted for with several models, including dark matter such as a dark photon $A'$ or an Axion-like particle~\cite{Chen:2015vqy}, supersymmetry~\cite{Kowalska:2015zja}, extra dimensions~\cite{Appelquist:2001jz}, additional Higgs bosons, or something completely new. 

\section{Fermilab Muon g-2 Experiment}

The BNL g-2 experiment E821 has left us with a 3.5$\sigma$ discrepancy with the standard model. The new experiment at Fermilab, E989, will measure 21 times the number of muons, which together with improvements in systematic uncertainty and expected theoretical improvements, could provide a $>7\sigma$ discrepancy from the Standard Model. The expected precision before and after this experiment is shown in Table~\ref{tab:comp}. The increase in statistics will also allow for an unprecedented measurement of the muon EDM, which will provide a two order of magnitude improvement in precision over the previous best measurement~\cite{Chislett:2016jau}. Complete details of the Muon g-2 experiment are given in Ref.~\cite{Grange:2015fou}.

To measure the anomalous magnetic moment, polarized muons are injected into a superconducting magnetic storage ring, and the muons will precess in the magnetic field. Polarized muons are produced naturally from pion decay and injected into a storage ring with a uniform magnetic field and cyclotron frequency given by
\begin{equation} 
\omega_c = \frac{e}{m\gamma}B
\label{eq:cycl}
\end{equation}
The spin precession frequency of the muons is
\begin{equation} 
\omega_s = \frac{e}{m\gamma}B(1+\gamma a_\mu),
\label{eq:spin}
\end{equation}
and we measure the difference between the two:
\begin{equation}
\omega_a = \omega_s - \omega_c = \frac{e}{m}a_\mu B
\label{eq:diff}
\end{equation}

Measurements of $\omega_a$ and $B$ provide $a_\mu$, which is related to the frequency $\omega_a$ and the magnetic field $\vec B$ by the relation
\begin{equation}
\vec\omega_a = -\frac{Qe}{m}[a_\mu \vec B - (a_\mu - (\frac{mc}{p})^2)\frac{\vec \beta \times \vec E}{c}]
\label{eq:omega_full}
\end{equation}
We perform the measurement at a magic momentum of $\gamma = 29.3$, so the electric field term cancels, leaving the relation
\begin{equation}
\vec\omega_a = -\frac{Qe}{m} a_\mu \vec B,
\label{eq:omega}
\end{equation}
which means that to measure $a_\mu$, we must precisely measure both $\omega_a$ and $\vec B$.

\begin{figure}
\centering
\subfloat[Template fit to double pulse.]{{
\includegraphics[width=6cm]{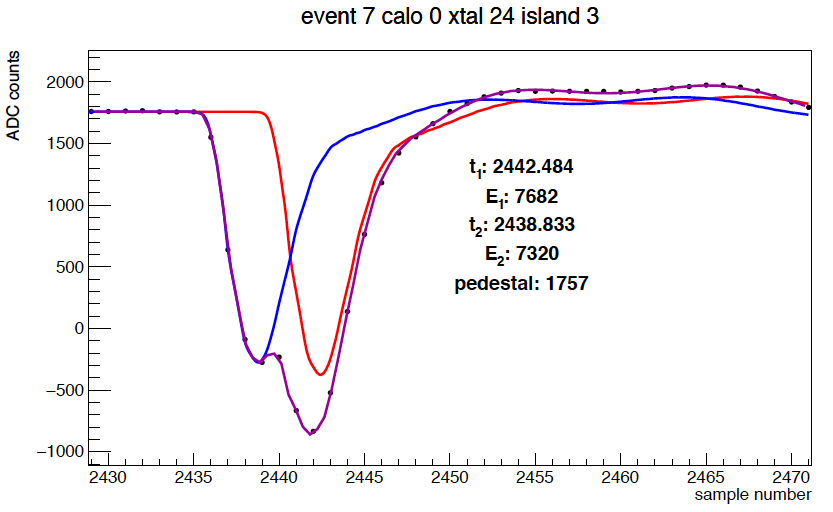}}}
\qquad
\subfloat[Five-parameter fit to muon decay signal.]{{
\includegraphics[width=6cm]{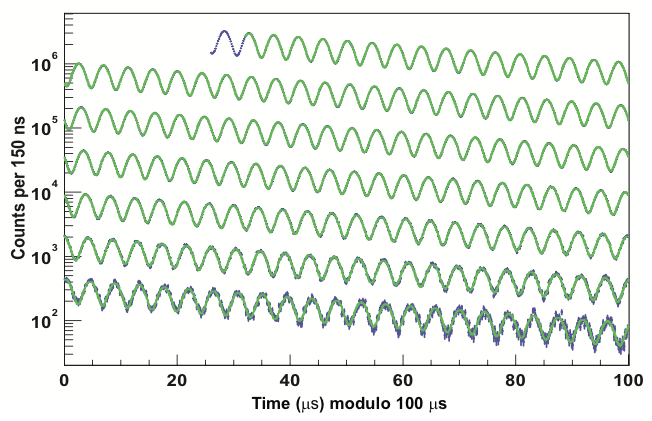}}}
\caption{a) Template fit to double pulse in calorimeter to resolve pileup. b) E821 five-parameter fit to determine precession frequency.}
\label{fig:wiggle}
\end{figure}

The $\omega_a$ measurement is performed by detecting positrons  from muon decay and fitting the time distribution of the decays with a five-parameter fit, and example of which is shown in Fig.~\ref{fig:wiggle}. The positrons are detected using 24 calorimeters each composed of 54 PbF$_2$ Cherenkov crystals~\cite{Fienberg:2014kka} read by silicon photomultipliers and recorded using custom 12 bit 800 MHz waveform digitizers~\cite{Chapelain:2015esj}, and read out using a GPU-based data acquisition system~\cite{Gohn:2015bla}. The calorimeters will be calibrated using a state-of-the-art laser calibration system~\cite{Anastasi:2015ssy}. Calibration runs will take place in the months before the accelerator complex is complete, but also in situ measurements will occur during data taking.

\begin{table}\centering
\begin{tabular}{|c|c|c|}
\hline
Uncertainty $\delta(a_{\mu})$ & Current value (ppb) & E989 Projection (ppb)\\
\hline
Theory & 420 & 310 \\
Experiment & 540 & 140\\
\hline
\end{tabular}
\caption{Current and expected precision on $a_\mu$ from theory and experiment.}
\label{tab:comp}
\end{table}


To make the measurement more precise, the precession frequency of the protons in the NMR probes $\omega_p$ is measured as a proxy for $\vec B$. The relationship between $\omega_a$ and $B$ then transforms to
\begin{equation}
\omega_a = \frac{eB}{m}a_\mu \rightarrow a_\mu = \frac{\omega_a / \omega_p}{\mu_\mu / \mu_p - \omega_a / \omega_p}
\label{eq:wp}
\end{equation}

The magnetic field in the storage ring will be 1.45 T at 5200 A, and must be constant in the muon storage region to $\pm0.5$ ppm. The field is homogenized by adding iron shims to remove quadrupole and sextapole asymmetries, adjusting metal plates on the top and bottom of the ring to change the effective $\mu$, and adding surface correction coils to add average field moments. 

\begin{figure}\centering
\includegraphics[width=13cm]{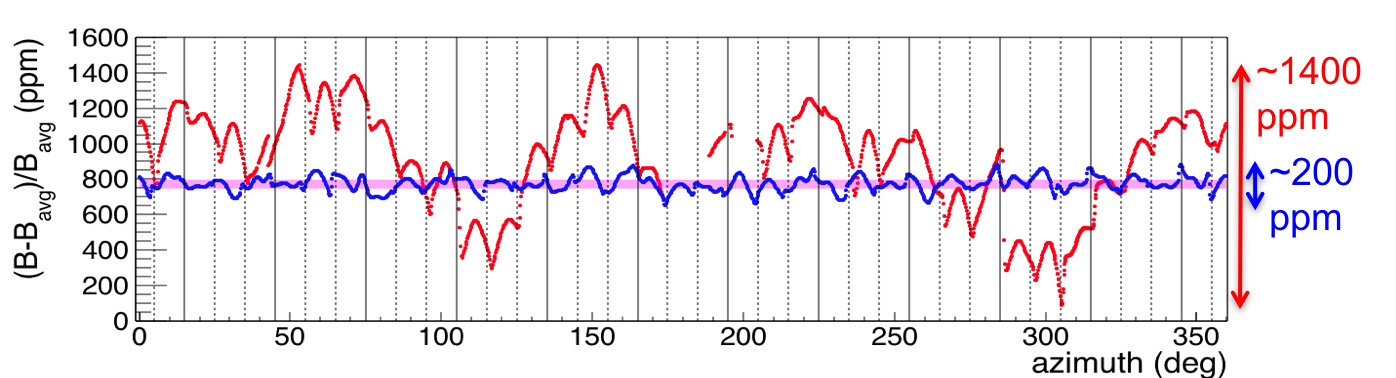}
\includegraphics[width=6cm]{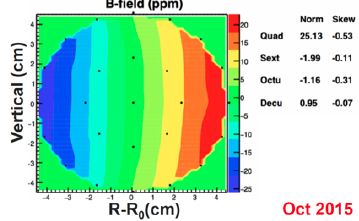}
\includegraphics[width=6cm]{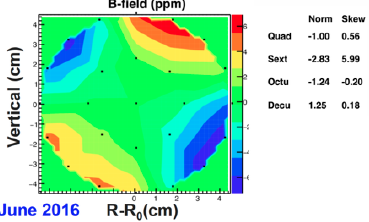}
\caption{Top: Azimuthal variation in magnetic field as of June 2016. The red curve indicates the field in October 2015, the blue shows the field variations in June 2016, and the width of the pink band is the desired variation, which was achieved in August, 2016. Bottom: Azimuthal average of magnetic field in October 2015 (left) and June 2016 (right).}
\label{fig:field}
\end{figure}

Fixed NMR probes measure time variations of the field during data taking. A trolley with mounted NMR probes periodically circumnavigates the interior of the ring to perform precision measurements of the field in the muon storage region, performing 6000 magnetic field measurements per trolly run. Probes are calibrated to provide measurement to 35 ppb using a 1.45 T MRI magnet. Shimming of the magnet has recently been completed. To perform the shimming, a special trolly outfitted with 25 NRM probes measured the field inside the ring while being tracked with a laser tracking system. The shimming procedure was iterated until a azimuthal variation of 50 ppm was reached, which was mandated by our systematic error budget. The azimuthal dependence of the field as of June 2016 and the azimuthal average over a cross section of the ring are shown in Fig.~\ref{fig:field}.

\begin{figure}[t]
\centering
\subfloat[$\omega_a$]{{
\includegraphics[width=6cm]{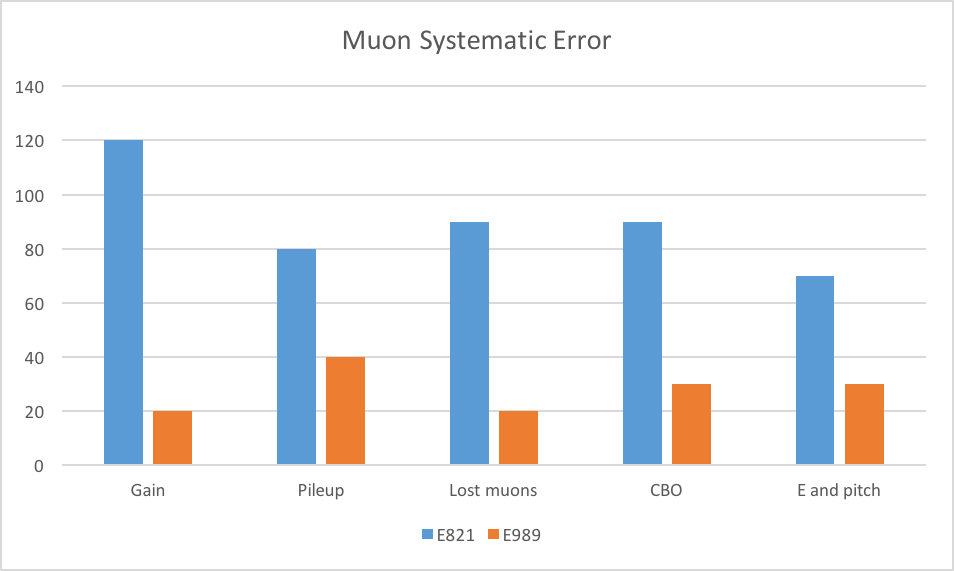}}}
\subfloat[Magnetic field]{{
\includegraphics[width=6cm]{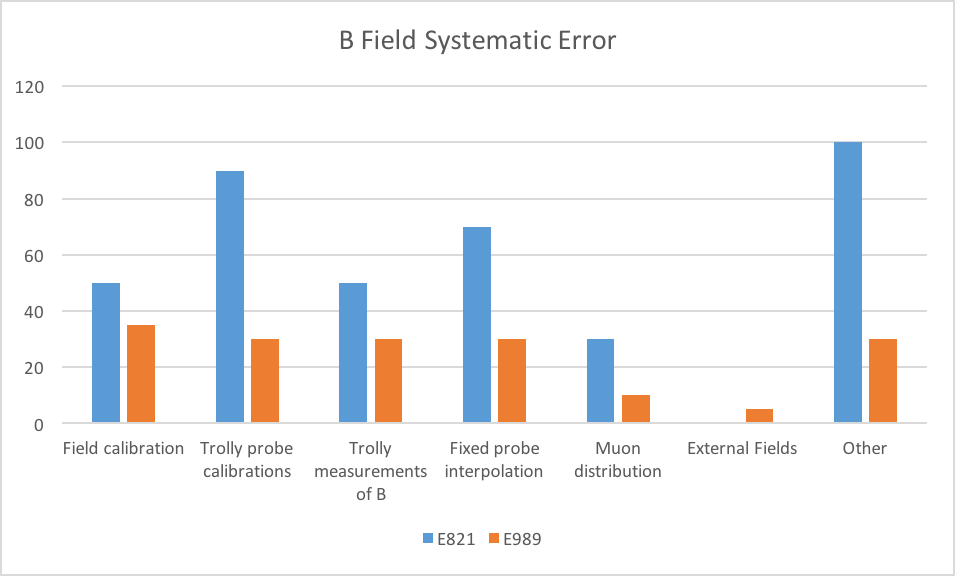}}}
\caption{Systematic error improvements to $\omega_a$ (left) and magnetic field (right) expected in E989.}
\label{fig:syst}
\end{figure}

The E989 experiment will collect 21 times the BNL statistics, which will reduce our statistical uncertainty by a factor of four, so it is necessary to reduce the systematic uncertainties by the same amount. Improved accelerator facilities will reduce beam power, have a $p_\pi$ closer to magic momentum, utilize a longer decay channel, and increase injection efficiency. Systematic uncertainties on $\omega_a$ will be decreased from 180 ppb in E821 to 70 ppb in E989 by using an improved laser calibration, a segmented calorimeter, better collimator in the ring, and improved tracker. Systematic uncertainties on $\omega_p$ will be decreased from 170 ppb in E821 to 70 ppb in E989 by improving the uniformity and monitoring of the magnetic field, increasing accuracy of position determination of trolly, better temperature stability of the magnet, and providing active feedback to external fields. Fig.~\ref{fig:syst} show a comparison of the expected sources of systematic uncertainty between the Brookhaven and Fermilab experiments. 

\section{Conclusion}

The new Muon g-2 experiment at Fermilab will measure the anomalous magnetic moment of the muon to 4 $\times$ the precision of the previous BNL measurement. If the previously measured value holds, this could provide a $>7\sigma$ discrepancy from the standard model, which would be a clear indication of new physics.  Homogenization of the ring's magnetic field is complete, and installation of detectors is beginning. The full system of detectors with GPU data acquisition and laser calibration will begin commissioning soon, and will be ready for the first injection of muons in Spring of 2017. Data taking will run through 2020.

\section*{References}
\bibliographystyle{h-physrev}
\bibliography{nufact}

\begin{thebibliography}{10}

\bibitem{Bennett:2006fi}
G.W. Bennett et~al.
\newblock {Final Report of the Muon E821 Anomalous Magnetic Moment Measurement
  at BNL}.
\newblock {\em Phys.Rev.}, D73:072003, 2006.

\bibitem{Dirac:1928hu}
Paul~A.M. Dirac.
\newblock {The Quantum theory of electron}.
\newblock {\em Proc.Roy.Soc.Lond.}, A117:610--624, 1928.

\bibitem{Nafe:1947zz}
J.~E. Nafe, E.~B. Nelson, and I.~I. Rabi.
\newblock {The Hyperfine Structure of Atomic Hydrogen and Deuterium}.
\newblock {\em Phys. Rev.}, 71:914--915, 1947.

\bibitem{Blum:2013xva}
Thomas Blum, Achim Denig, Ivan Logashenko, Eduardo de~Rafael, B.~Lee~Roberts,
  et~al.
\newblock {The Muon (g-2) Theory Value: Present and Future}.
\newblock 2013.

\bibitem{Aoyama:2012wk}
Tatsumi Aoyama, Masashi Hayakawa, Toichiro Kinoshita, and Makiko Nio.
\newblock {Complete Tenth-Order QED Contribution to the Muon g-2}.
\newblock {\em Phys.Rev.Lett.}, 109:111808, 2012.

\bibitem{Gnendiger:2013pva}
C.~Gnendiger, D.~Stöckinger, and H.~Stöckinger-Kim.
\newblock {The electroweak contributions to $(g-2)_\mu$ after the Higgs boson
  mass measurement}.
\newblock {\em Phys. Rev.}, D88:053005, 2013.

\bibitem{Davier:2010nc}
Michel Davier, Andreas Hoecker, Bogdan Malaescu, and Zhiqing Zhang.
\newblock {Reevaluation of the Hadronic Contributions to the Muon g-2 and to
  $\alpha(MZ)$}.
\newblock {\em Eur.Phys.J.}, C71:1515, 2011.

\bibitem{Hagiwara:2011af}
Kaoru Hagiwara, Ruofan Liao, Alan~D. Martin, Daisuke Nomura, and Thomas
  Teubner.
\newblock {$(g-2)_\mu$ and $\alpha(M_Z^2)$ re-evaluated using new precise
  data}.
\newblock {\em J. Phys.}, G38:085003, 2011.

\bibitem{Ablikim:2015orh}
M.~Ablikim et~al.
\newblock {Measurement of the $e^+ e^− \to \pi^+ \pi^−$ cross section
  between 600 and 900 MeV using initial state radiation}.
\newblock {\em Phys. Lett.}, B753:629--638, 2016.

\bibitem{Lees:2012cj}
J.~P. Lees et~al.
\newblock {Precise Measurement of the $e^+ e^- \to \pi^+\pi^- (\gamma)$ Cross
  Section with the Initial-State Radiation Method at BABAR}.
\newblock {\em Phys. Rev.}, D86:032013, 2012.

\bibitem{Babusci:2012rp}
D.~Babusci et~al.
\newblock {Precision measurement of $\sigma(e^+e^-\rightarrow
  \pi^+\pi^-\gamma)/ \sigma(e^+e^-\rightarrow \mu^+\mu^-\gamma)$ and
  determination of the $\pi^+\pi^-$ contribution to the muon anomaly with the
  KLOE detector}.
\newblock {\em Phys. Lett.}, B720:336--343, 2013.

\bibitem{Blum:2015you}
T.~Blum, P.~A. Boyle, T.~Izubuchi, L.~Jin, A.~Jüttner, C.~Lehner, K.~Maltman,
  M.~Marinkovic, A.~Portelli, and M.~Spraggs.
\newblock {Calculation of the hadronic vacuum polarization disconnected
  contribution to the muon anomalous magnetic moment}.
\newblock {\em Phys. Rev. Lett.}, 116(23):232002, 2016.

\bibitem{Blum:2016xpd}
T.~Blum et~al.
\newblock {Lattice calculation of the leading strange quark-connected
  contribution to the muon $g − 2$}.
\newblock {\em JHEP}, 04:063, 2016.

\bibitem{Blum:2015gfa}
Thomas Blum, Norman Christ, Masashi Hayakawa, Taku Izubuchi, Luchang Jin, and
  Christoph Lehner.
\newblock {Lattice Calculation of Hadronic Light-by-Light Contribution to the
  Muon Anomalous Magnetic Moment}.
\newblock {\em Phys. Rev.}, D93(1):014503, 2016.

\bibitem{Chen:2015vqy}
Chien-Yi Chen, Hooman Davoudiasl, William~J. Marciano, and Cen Zhang.
\newblock {Implications of a light “dark Higgs” solution to the $g_μ$-2
  discrepancy}.
\newblock {\em Phys. Rev.}, D93(3):035006, 2016.

\bibitem{Kowalska:2015zja}
Kamila Kowalska, Leszek Roszkowski, Enrico~Maria Sessolo, and Andrew~J.
  Williams.
\newblock {GUT-inspired SUSY and the muon g − 2 anomaly: prospects for LHC 14
  TeV}.
\newblock {\em JHEP}, 06:020, 2015.

\bibitem{Appelquist:2001jz}
Thomas Appelquist and Bogdan~A. Dobrescu.
\newblock {Universal extra dimensions and the muon magnetic moment}.
\newblock {\em Phys. Lett.}, B516:85--91, 2001.

\bibitem{Chislett:2016jau}
Rebecca Chislett.
\newblock {The muon EDM in the g-2 experiment at Fermilab}.
\newblock {\em EPJ Web Conf.}, 118:01005, 2016.

\bibitem{Grange:2015fou}
J.~Grange et~al.
\newblock {Muon (g-2) Technical Design Report}.
\newblock 2015.

\bibitem{Fienberg:2014kka}
A.T. Fienberg, L.P. Alonzi, A.~Anastasi, R.~Bjorkquist, D.~Cauz, et~al.
\newblock {Studies of an array of PbF$_2$ Cherenkov crystals with large-area
  SiPM readout}.
\newblock {\em Nucl.Instrum.Meth.}, A783:12--21, 2015.

\bibitem{Chapelain:2015esj}
Antoine Chapelain.
\newblock {Development of the electromagnetic calorimeter waveform digitizers
  for the Fermilab Muon g-2 experiment}.
\newblock {\em PoS}, EPS-HEP2015:280, 2015.

\bibitem{Gohn:2015bla}
Wesley Gohn.
\newblock {Data Acquisition for the New Muon $g$-$2$ Experiment at Fermilab}.
\newblock {\em J. Phys. Conf. Ser.}, 664(8):082014, 2015.

\bibitem{Anastasi:2015ssy}
A.~Anastasi et~al.
\newblock {Test of candidate light distributors for the muon (g$-$2) laser
  calibration system}.
\newblock {\em Nucl.Instrum.Meth.}, A788:43--48, 2015.

\end{thebibliography}

\end{document}